\documentclass[5p]{elsarticle}

\usepackage{graphicx}
\usepackage{amsmath,amssymb,amsfonts}
\usepackage{url}

\begin{document}

\begin{frontmatter}

  \title{A Redistribution Tool for Long-Term Archive of Astronomical Observation Data\tnoteref{t1}}

  \tnotetext[t1]{The source code has been upload on gitee: \url{https://gitee.com/AstroTJU/AstroLayout}.}

  \author[1,3]{Chao Sun}
  \ead{sch@tju.edu.cn}

  \author[1,3]{Ce Yu\corref{cor1}}
  \ead{yuce@tju.edu.cn}

  \author[2,3]{Chenzhou Cui}
  \ead{ccz@bao.ac.cn}

  \author[2,3]{Boliang He}
  \ead{hebl@nao.cas.cn}

  \author[1,3]{Jian Xiao}
  \ead{xiaojian@tju.edu.cn}

  \author[1,3]{Zhen Li}
  \ead{lizhencs@tju.edu.cn}

  \author[1,3]{Shanjiang Tang}
  \ead{tashj@tju.edu.cn}

  \author[1,3]{Jizhou Sun}
  \ead{jzsun@tju.edu.cn}

  \cortext[cor1]{Corresponding author}

  \address[1]{College of Intelligence and Computing, Tianjin University,
  	135 Yaguan Road Jinnan District, Tianjin 300350, China}
  \address[2]{National Astronomical Observatories, Chinese Academy of Sciences,
    20A Datun Road, Chaoyang District, Beijing 100012, China}
  \address[3]{NAOC-TJU Joint Research Center in Astro-Informatics,
    135 Yaguan Road Jinnan District, Tianjin 300350, China}

  \begin{abstract}
  Astronomical observation data require long-term preservation,
  and the rapid accumulation of observation data
  makes it necessary to consider the cost of long-term archive storage.
  In addition to low-speed disk-based online storage, optical disk or tape-based offline storage
  can be used to save costs.
  However, for astronomical research that requires historical data (particularly time-domain astronomy),
  the performance and energy consumption of data-accessing techniques cause problems
  because the requested data (which are organized according to observation time)
  may be located across multiple storage devices.
  In this study, we design and develop a tool referred to as AstroLayout to
  redistribute the observation data using spatial aggregation.
  The core algorithm uses graph partitioning
  to generate an optimized data placement
  according to the original observation data statistics and the target storage system.
  For the given observation data,
  AstroLayout can copy the long-term archive
  in the target storage system in accordance with this placement.
  An efficiency evaluation shows that
  AstroLayout can reduce the number of devices activated when responding to data-access requests
  in time-domain astronomy research.
  In addition to improving the performance of data-accessing techniques,
  AstroLayout can also reduce the storage system’s power consumption.
  For enhanced adaptability, it supports storage systems of any media,
  including optical disks, tapes, and hard disks.
  \end{abstract}


\end{frontmatter}

\section{Introduction}

Astronomy is one of the few sciences that rely entirely on observations;
however, in contrast to other such sciences, the observational data of celestial bodies cannot be reproduced.
Thus, long-term preservation of the data is necessary,
and a rapid accumulation of data will lead to increasingly high archiving costs
\citep{Layne2012LongTerm} \citep{Stoehr2015ALMA}.
To reduce these costs,
optical disk or tape-based offline storages, as well as low-speed disk-based online storage,
can be used for archiving \citep{Wan2014Optical}.
The devices that realize these low-cost storage systems are always offline when not in use
because most of the data are requested only occasionally.

\begin{figure}[ht]
  \centering
  \includegraphics{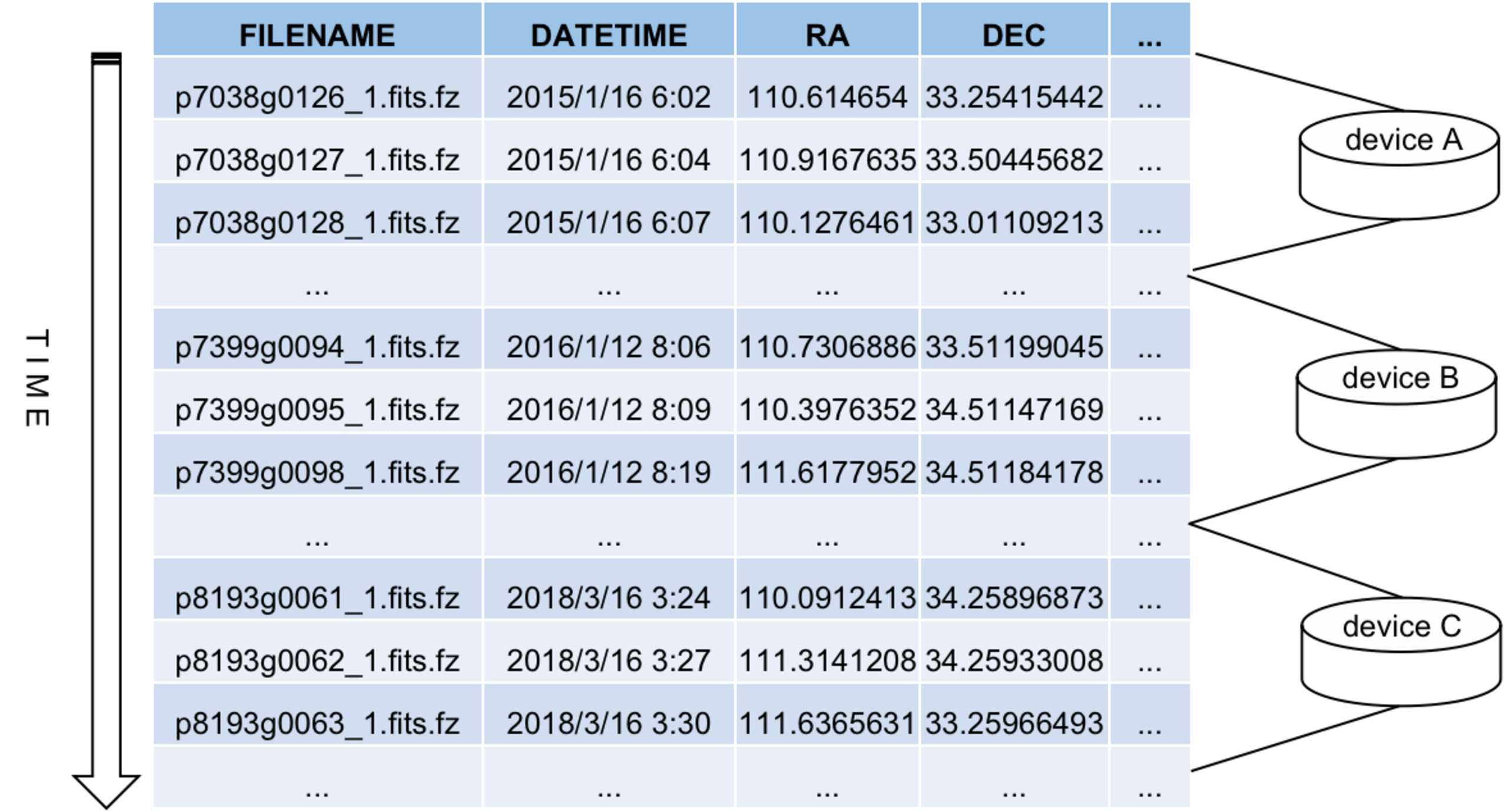}
  \caption{Traditional data placement.
  The flexible image transport system (FITS) files are organized according to observation time;
  as such, files labeled with similar observation times are likely to be stored in the same device,
  whereas the data of specific observation regions may be distributed over multiple devices.}
  \label{fig:fig1}
\end{figure}

However, the data are generally organized according to observation time;
this is the traditional data-placement method (see Figure~\ref{fig:fig1}).
For astronomical research based on historical data (particularly in time-domain astronomy),
the research problem often concerns on the changes of celestial objects over a period of time.
Therefore, the data requests are mainly for a specific observation region,
and the requested data are often located across multiple devices in the storage system.
Consequently, multiple devices must be activated
to serve a data-access request for a specific observation region;
this consumes more time and energy, which is detrimental to
the performance and energy efficiency of the data-accessing procedures.

\begin{figure}[ht]
  \centering
  \includegraphics{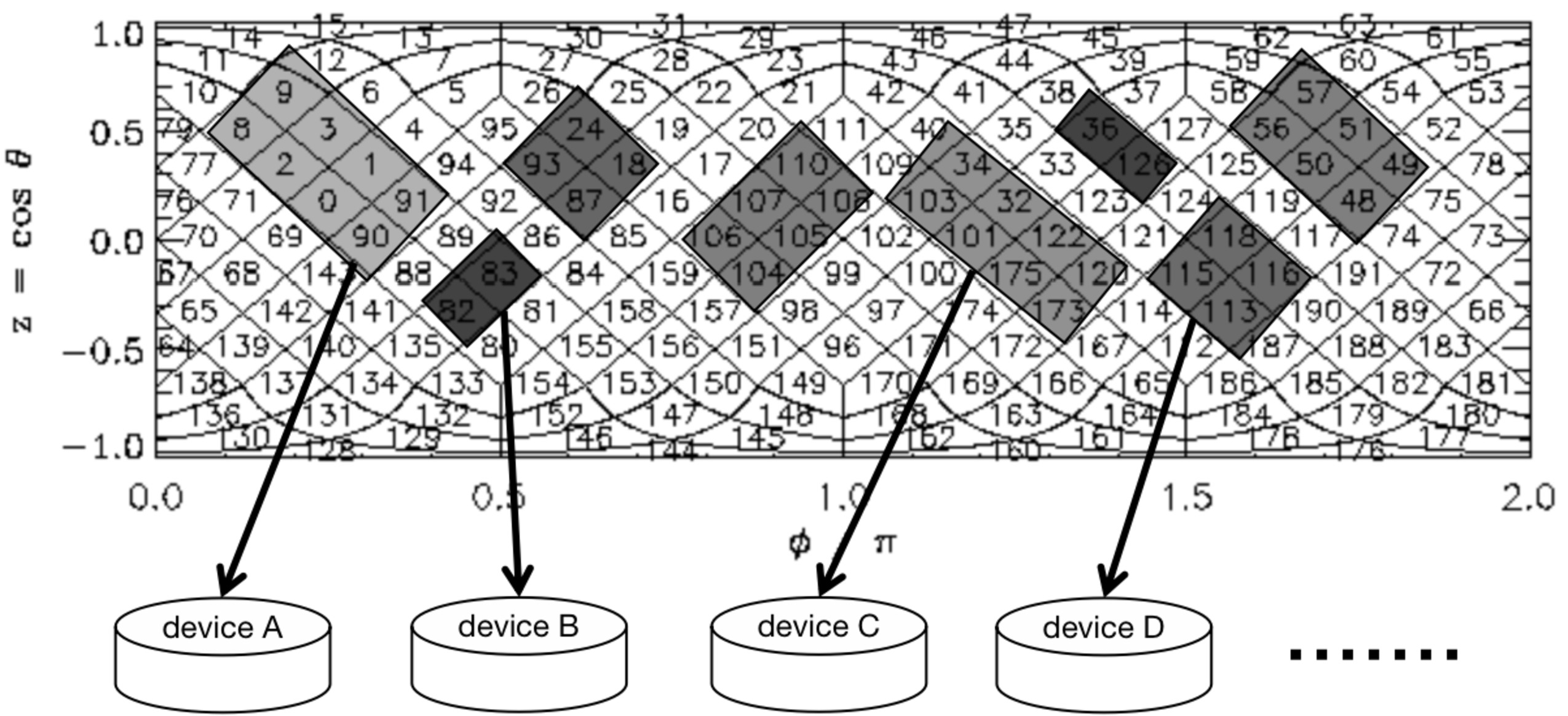}
  \caption{Spatial aggregation data placement (SADP).
  Data from neighboring observation regions are stored in the same device,
  and the color shading indicates the data density.}
  \label{fig:fig2}
\end{figure}

Using a different data-placement procedure
(one that aggregates the stored data according to spatial attributes)
can significantly reduce the number of devices activated
while also improving the system's performance and energy efficiency.
Moreover, it prolongs the lifetimes of the devices,
especially for disk-based storage systems.
In this work,
we design and develop a tool called AstroLayout to redistribute the astronomical data
using spatial aggregation; this was proposed in our previous work \citep{Li2018Gpdl}.
AstroLayout can compute an optimized data placement
(referred to as a spatial aggregation data placement, SADP),
in which the data of neighboring observation regions are aggregated into one group
(see Figure~\ref{fig:fig2}).
Given a set of flexible image transport system (FITS) files,
AstroLayout can generate a SADP using graph partitioning
and then archive the files to the target storage devices according to the generated data placement.
The target storage system can consist of any type of storage media,
including optical disks, tapes, and hard disks.
Moreover, AstroLayout supports several additional features for robustness and reliability.

The remainder of this paper is structured as follows:
Section 2 introduces the related work,
Section 3 details the process and the structure of AstroLayout,
Section 4 contains a discussion of the evaluation,
and the last section concludes our work.

\section{Related Work}

AstroLayout is a tool to redistribute the long-term archives of observation data
in low-cost storage system;
its purpose is to improve the performance and the energy efficiency of the system
when responding to data-access requests for specific observation regions.
The core algorithm of AstroLayout computes an optimized data placement
using spatial aggregation;
this aggregates the FITS files of neighboring observation regions into groups
and stores it into devices of target storage.

\subsection*{Data-Placement Approaches.}

Many studies have been conducted on data placement to improve the performance of storage systems;
however, most of these have focused on a distributed environment \citep{Li2016EStore}
or even heterogeneous environments \citep{Gao2018Storage} \citep{Ambore2019Survey}.
Only a few studies have considered the optimization of low-cost storage,
and the general solution is to aggregate the correlated data together.
Some researchers have identified the correlated data using historical data access \citep{Hu2018Aggregating},
whereas others used analysis of the data attributes \citep{Yan2017DataLayout}.
These approaches are preferable to real-time service systems;
however, they are not suitable for the long-term archive in which the data are rarely accessed.
This archive does not need to introduce a cache or duplicated data.

\subsection*{Spherical Surface Partition.}

\begin{figure}[ht]
	\centering
	\includegraphics{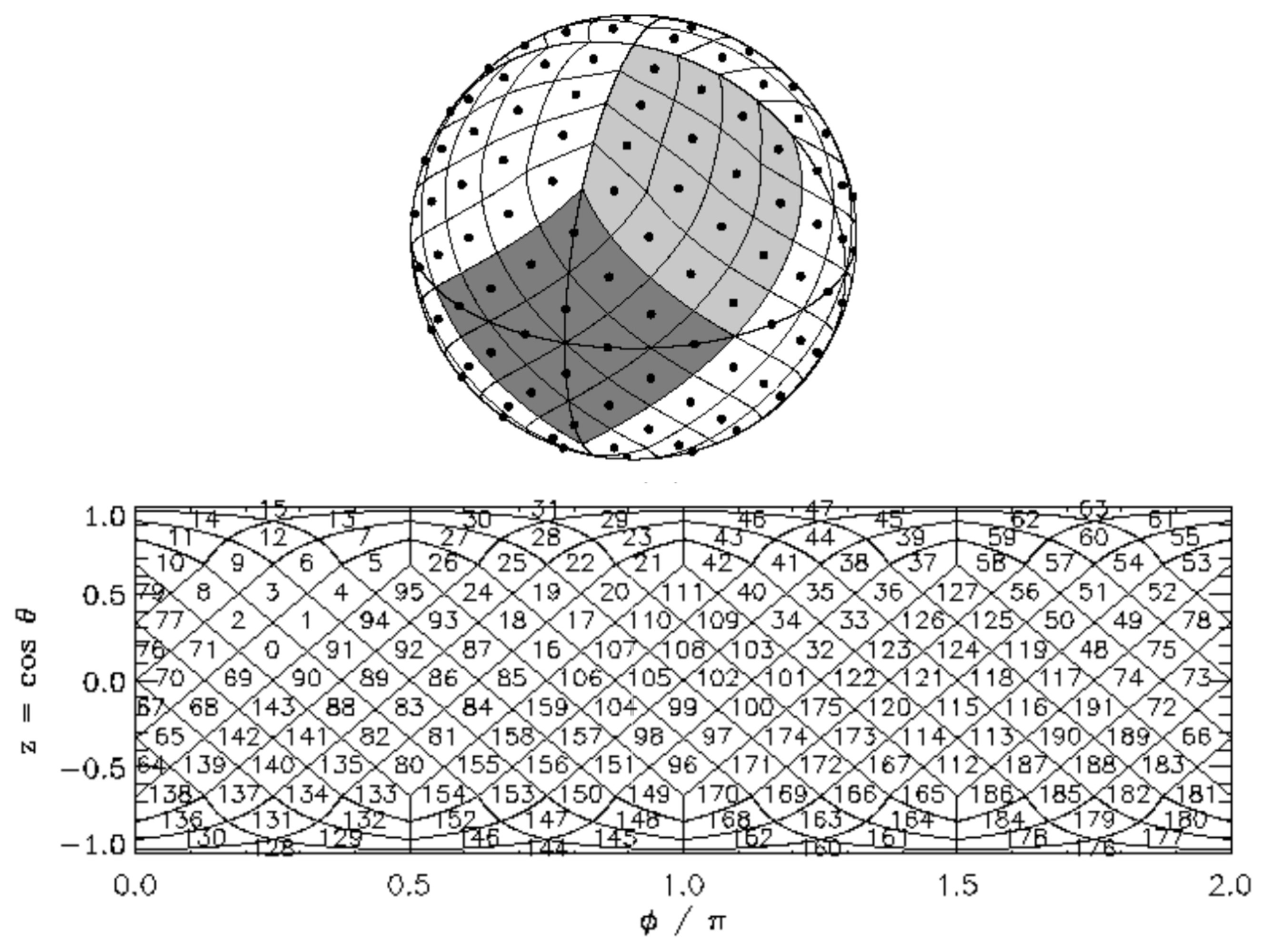}
	\caption{Healpix partitioning and the NESTED numbering scheme ($NSIDE = 4$).}
  \label{fig:fig3}
\end{figure}

Grouping FITS files by observation region is one of the key components of this method,
and the spherical surface partition methods are suitable for this type of data.
A hierarchical triangular mesh (HTM) is a multi-level, recursive decomposition of a sphere \citep{HTM2010},
while the hierarchical equal area isolatitude pixelization of a sphere
(HEALPix) partitions a sphere surface into cells of equal surface area \citep{HEALPix2005};
both of these partitions are widely used for spherical surfaces.
In contrast with HTM, HEALPix was originally designed for spherical spaces,
applying equal cell areas over the entire spherical surface;
thus, the distances between a cell and each of its neighbors can be weighted equally.
For this reason, AstroLayout uses HEALPix as the spherical surface partitioning method.
HEALPix supports two different pixel numbering schemes,
and in the NESTED scheme the pixel number grows with consecutive hierarchical subdivisions
on a tree structure seeded by the twelve base-resolution pixels.
Figure~\ref{fig:fig3} shows the HEALPix partitioning and the NESTED numbering scheme.

\subsubsection*{Data Aggregation Method.}

Aggregating the HEALPix cells of FITS files into groups is similar to graph partitioning,
which is used to solve optimization problems arising in numerous research fields.

\begin{figure}[ht]
	\centering
	\includegraphics{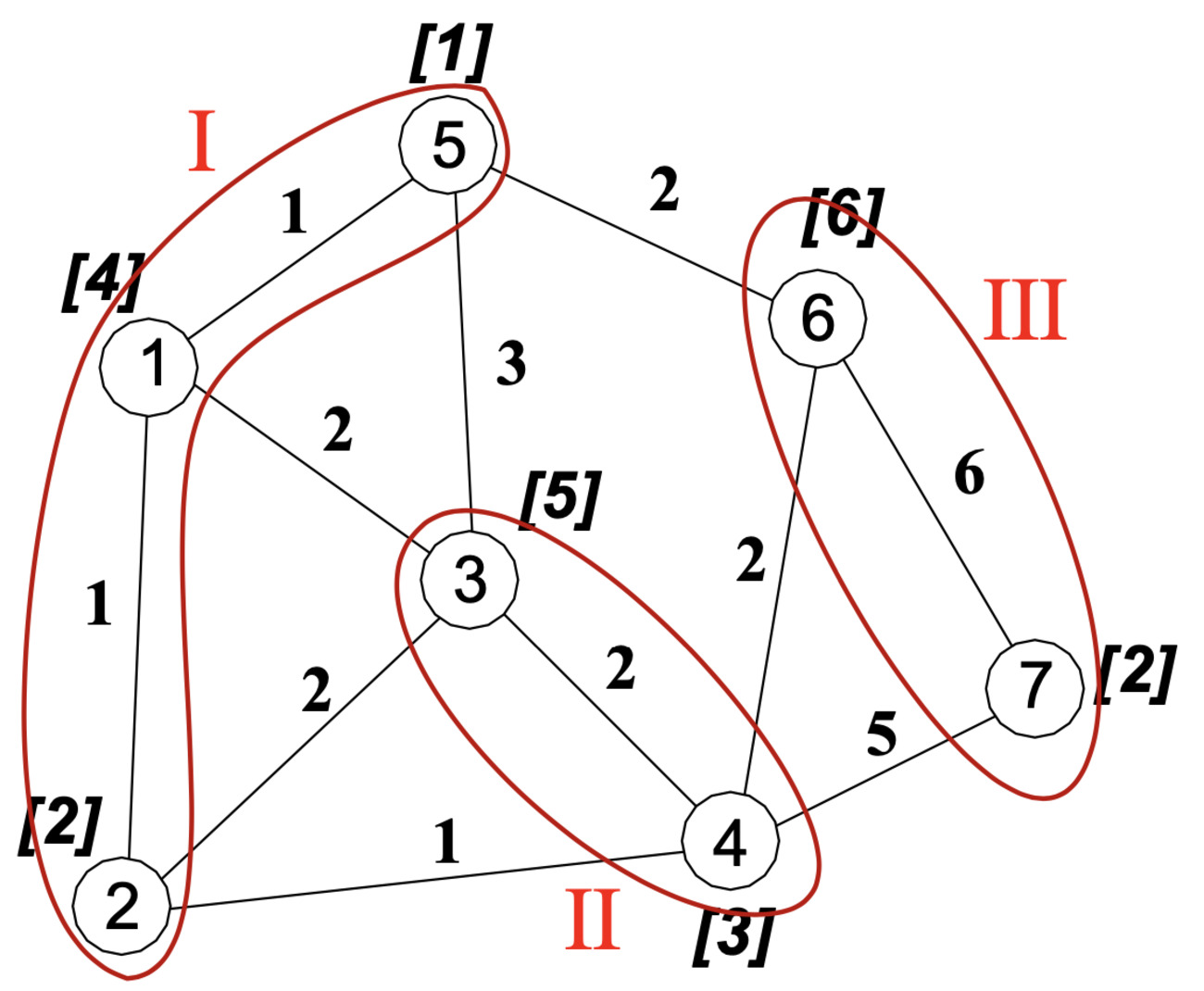}
	\caption{Sample of 3-way graph partitioning.}
  \label{fig:fig4}
\end{figure}

Given a graph $G = (V, E)$ with $V$ vertices and $E$ edges,
graph partitioning is the problem of how to partition $G$ into
smaller sub-graphs with specific properties.
In a $k$-way partitioning (see Figure~\ref{fig:fig4}), the aim is to
divide the vertices into $k$ smaller parts of roughly equal size, while
minimizing the weight of the edges between the separated $k$ sub-graphs
\citep{Andreev2006Balanced}.
For a weighted graph $G = (V, E)$, in which every vertex and edge has weights,
a good partitioning is satisfied by performing the following:
\begin{enumerate}[1.]
  \item minimizing the summed weights of edges between the separated sub-graphs,
  \item roughly equalizing the summed vertex weights of different sub-graphs.
\end{enumerate}

Because graph partitioning is a difficult problem, the solutions are based upon heuristics,
using two broad categories:
one that works locally and one that considers global connectivity.
To solve this difficult graph partitioning problem, AstroLayout uses METIS \citep{Metis1999},
a popular software package that partitions graphs by considering global connectivity.

\section{AstroLayout}

AstroLayout is a tool to aggregate the FITS files of neighboring observation regions.
The core algorithm of AstroLayout uses graph partitioning
and calculates an optimized data placement solution using spatial aggregation.
AstroLayout outputs a new distribution using SADP for long-term observation data
in the target storage system;
this system can be realized in any type of storage media,
including optical disk, tape, and hard disk.

AstroLayout also supports several additional features for robustness and reliability.
It can serialize the status of the program during operation and
restore the running progress even if a power failure unexpectedly occurs.
Furthermore, a logging module can record the information during the program's operation,
including the execution logs, the error messages, and distribution progress information.

\subsection{AstroLayout Workflow}

\begin{figure}[ht]
	\centering
	\includegraphics{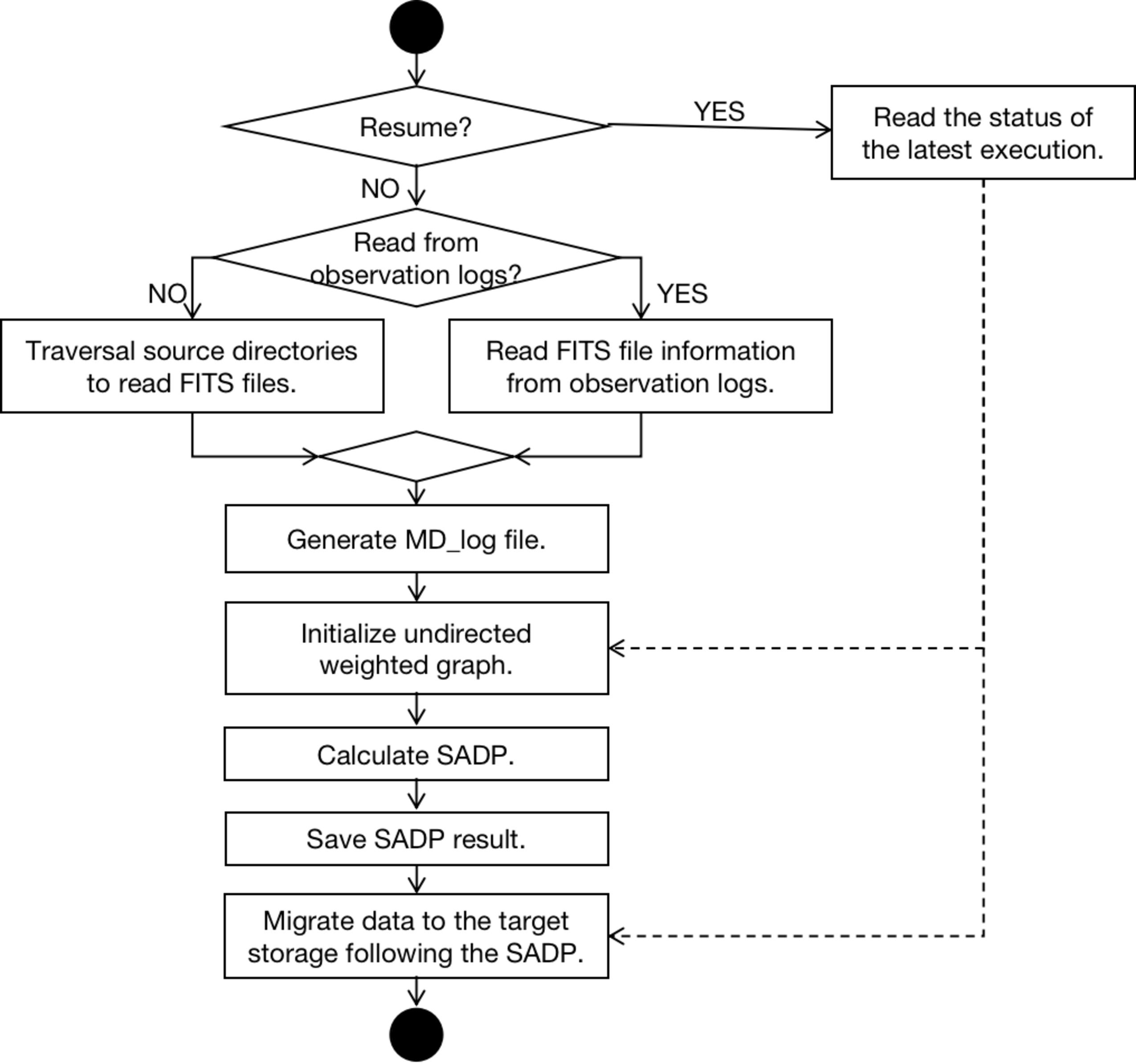}
	\caption{Workflow of AstroLayout.}
  \label{fig:fig5}
\end{figure}

The schematic workflow of AstroLayout is shown in Figure~\ref{fig:fig5}.
To begin, the resuming feature can be triggered by a command line parameter;
this checks the status of the latest execution to determine which procedure should be implemented.

For the main procedure, AstroLayout operates as follows:
\begin{enumerate}[1.]
  \item It gathers the FITS file information, either directly from the files
  or from the observation logs.
  \begin{enumerate}[a.]
    \item To read the FITS files directly,
    it extracts their center right ascension and declination (RA/Dec) values.
    Furthermore, it gathers information about the files themselves, such as their paths and sizes.
    \item Reading from observation logs can significantly improve the processing efficiency;
    however, the observation logs from the observation systems
    may contain insufficient information about the FITS files
    --- for instance, they might not contain the file sizes.
    In such cases, the specified value from the configuration file are used.
  \end{enumerate}
  \item It converts the gathered spatial information (one for each file)
  into the HEALPix pixel numbering system
  and caches the information into MD\_log file for resuming
  (including the file location, the file size, the spatial information, and the HEALPix pixel corresponding).
  \item An undirected weighted graph is generated for these pixels,
  using the total file size of the corresponding FITS files as its vertex weights.
  The edge weights corresponding to the distances between pixels.
  \item Given the maximum capacity of the target storage devices,
  AstroLayout attempts to partition this graph into several balanced sub-graphs,
  such that a SADP can be generated from the result.
  \item It saves the SADP solution for resuming.
  \item To archive the FITS files,
  a target storage with a specified number of devices needs to be prepared.
  AstroLayout distributes the FITS files in the storage following the generated SADP.
  This step takes a long time, depending on the size of the dataset.
\end{enumerate}

During execution, AstroLayout serializes the status of the program;
this can be used for resuming operations.
It also records the program information,
including the execution logs, error messages, and distribution progress.

\subsection{AstroLayout Architecture}

\begin{figure}[ht]
  \centering
  \includegraphics{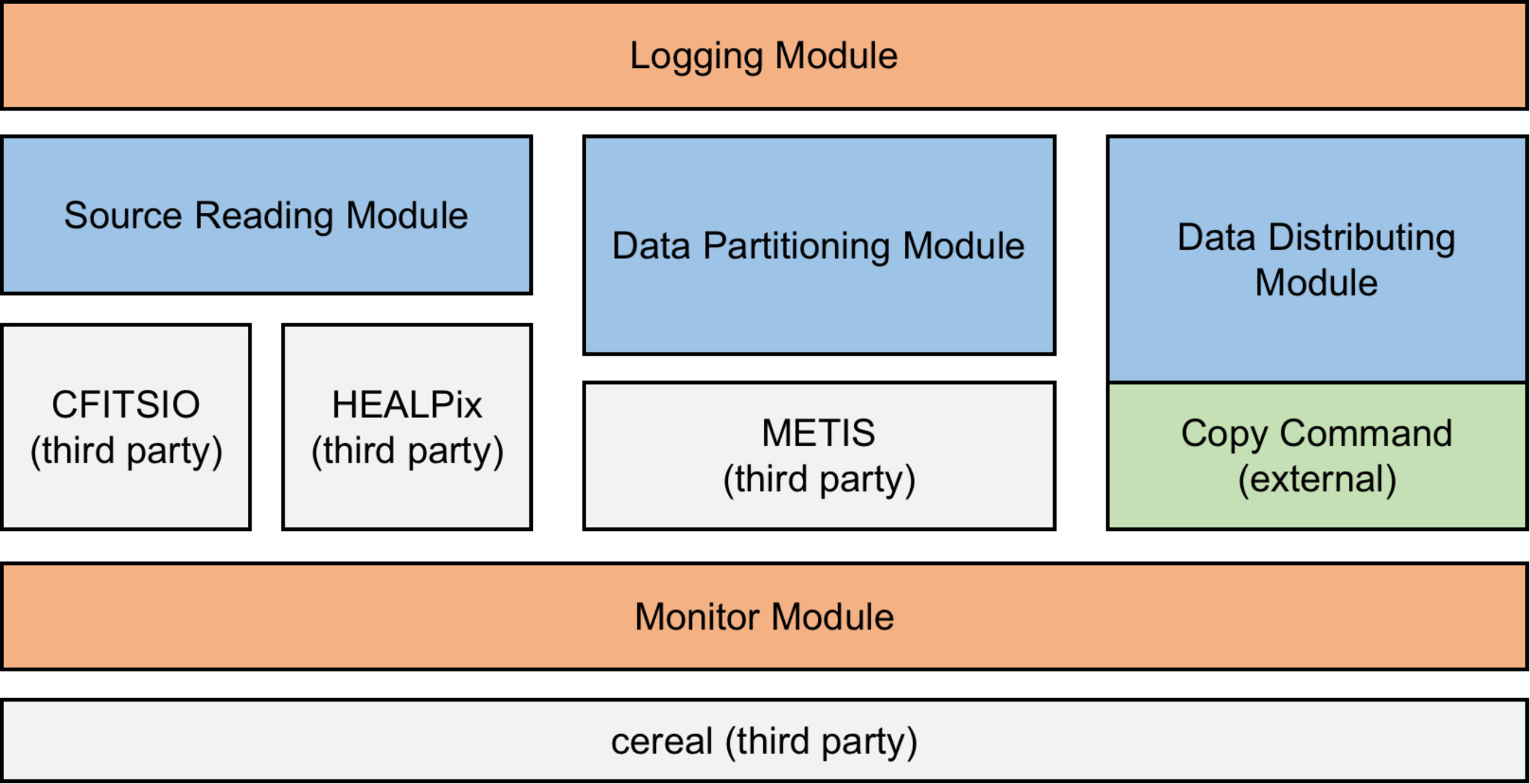}
  \caption{Architecture of AstroLayout.}
  \label{fig:fig6}
\end{figure}

Figure \ref{fig:fig6} illustrates the architecture of AstroLayout
it contains three main modules and two auxiliary modules.
The main modules are the source-reading module, data-partitioning module,
and data-distributing module;
all these modules interact with the auxiliary modules:
the logging module and the monitor module.
Besides this, the program uses several third-party development libraries:
\begin{enumerate}[1.]
  \item CFITSIO \citep{Cfitsio1999}: a FITS file subroutine library,
  \item HEALPix \citep{HEALPix2005}: hierarchical equal area isolatitude pixelization,
  \item METIS \citep{Metis1999}: serial graph partitioning and fill-reducing matrix ordering,
  \item cereal \citep{Cereal2017}: a C++11 library for serialization.
\end{enumerate}
These libraries are used for implementing specific functions,
and can be replaced with other equivalent libraries.

\subsubsection{Source-Reading Module}

In the source-reading module, AstroLayout gathers information about the FITS files,
and statistically generates a data quantity to be allocated to each HEALPix pixel;
this represents the data distribution on the celestial sphere surface.

The file information can be retrieved via two methods:
the first scans the FITS files directly from the source directories directly,
the second extracts information from the observation logs.
The former must directly reading for all the source files;
the latter is more efficient but may miss the the file sizes.
For sources containing similar-sized files, it is convenient to use the latter method.

To calculate the HEALPix pixels of FITS files,
CFITSIO is used to parse the header information and extract the values
of RA/Dec, according to the keywords listed in the configuration file;
then, it converts these values into the HEALPix pixel, using HEALPix C++ subroutines.

\subsubsection{Data-Partitioning Module}

Given the data distribution on the celestial sphere surface,
the data-partitioning module generates a SADP,
using graph partitioning to aggregate the HEALPix pixels.

The purpose of AstroLayout is to
compute a relationship that maps FITS files to devices in a way that
satisfies the following conditions:
\begin{enumerate}[1.]
  \item \textbf{Spatial locality}:
  The FITS files for one region on the celestial sphere surface must be aggregated
  together, owning to the spatial locality of data requests in time-domain astronomy;
  that is, a single data-access request often focuses on a specific region over a period of time
  to enable researchers to study the changes of one or more celestial objects.
  \item \textbf{Load balancing}:
  For long-term preservation, the archives often hold static historical data.
  For a more optimal storage utilization, the total capacity of the archive storage
  should approximately equate to the total size of the FITS files;
  thus, every device must store as many FITS files as possible.
  In other words, the data distributed to all the devices should be optimally balanced.
\end{enumerate}
These can be converted into a graph partitioning problem and solved using METIS.
The conversion steps are as follows:
\begin{enumerate}[1.]
  \item
  Map HEALPix pixels to the vertices of the graph,
  and set the weight of each vertex as
  the size of the FITS files corresponding to the HEALPix pixel.
  \item
  Let the weight of edges between vertices correspond to
  the distance between the relevant HEALPix pixels.
  The closer the HEALPix pixels, the larger the weight.
  The weight is also influenced by the weights of the vertices.
\end{enumerate}

Using METIS, partition the graph into several sub-graphs
and confirm that the total size of FITS files for each sub-graphs does not exceed
the capacity of the target storage devices.
If this is not fulfilled, increase the number of sub-graphs and repeat the above process.

\subsubsection{Data-Distributing Module}

In the data-distributing module, an archive storage is required;
and AstroLayout copies FITS file from the source directories into the archive storage
using the generated SADP.
This process will be the most time-consuming.

A custom copy program can be implemented in place of the system's copy command
to improve the performance and reliability.
During the copying process, the module verifies the existence of the target file
and compares the file sizes of the source file.
If the results show that the files are identical, duplication is not required.

The archive storage can be realized by optical disks, tapes, or hard disks.
Files are stored in the devices one-by-one following the SADP.
If using optical disks and tapes, the devices must be exchanged during the copying process,
and the program will confirm the mounted device’s information to enhance usability.

\subsubsection{Auxiliary Modules}

The logging and monitor modules are designed to improve the robustness and reliability of AstroLayout.

The logging module records messages during the runtime;
these include the execution prompt, the error messages, and the distribution results for each FITS file.

The monitor module saves the running status during the program's operation
and contains all the running information.
When an emergency interruption halts the program (or in the event of an unexpected power failure),
it can restore the running progress when instructed.

\subsection{AstroLayout Usage}

The command line parameter used to trigger the resuming feature is \verb|--resume|.
When this parameter is used, AstroLayout checks the status from the status file
specified in the configuration file to determine which procedure to implement.

All runtime parameters must be prepared in a configuration file \textit{config.ini}
for AstroLayout.
This configuration file is in an initialization (INI) format,
an easy-to-read format widely used for configuration.
A demonstration can be found in the source directory and lists all the supported options.

Some of the configuration options are listed as follows.
\begin{enumerate}[1.]
  \item Option \verb|cereal| in \verb|[GLOBAL]| specifies the status file,
  to serialize the running status.
  \item Option \verb|dirs| in \verb|[SOURCE]| indicates the source directories.
  \item Option \verb|from_obs_log| in \verb|[SOURCE]| decides
  whether to gather the FITS file information directly from the files
  or from the observation logs.
  \item Options \verb|ra_keys| and \verb|dec_keys| in \verb|[FITS]| give
  the potential keywords of the center RA/Dec in the FITS files.
  \item Options in the section \verb|[OBSLOG]| define all the possible parameters
  for reading from observation logs.
  \item Option \verb|dirs| in \verb|[TARGET]| specifies the mount points of the devices.
  \item Option \verb|capacity| in \verb|[TARGET]| indicates
  the maximum capacities of the target storage devices.
  \item Option \verb|media| in \verb|[TARGET]| specifies the media type of the devices.
  If optical disks or tapes are used for the target storage,
  the mount points may be identical for all devices
  and the program will wait for a new device to become available.
  \item Option \verb|reserve_filepath| in \verb|[DISTRIBUTE]| indicates
  whether to preserve the source path structure when archiving.
\end{enumerate}

\section{Evaluation}

Our evaluation of AstroLayout is based on the observation data obtained by
the Antarctic survey telescopes (AST3) \citep{Cui2008Antarctic}.
This project generated 71589 FITS files in 2016; however, 22826 of them are focused on one area
(approximately 220~MB$+$ for each file).
In this experiment, we extracted the file information from the the official observation log
excluded the records for the fixed area,
and chose Seagate disks with the capacity of 1TB for the archive in the evaluation.

We simulated a request pool to verify the effectiveness of AstroLayout.
The request pool contained five groups of requests, querying data
in different scales of sky area
($1^\circ$, $2^\circ$, $3^\circ$, $4^\circ$ and $5^\circ$),
each group contains 1000 requests.
To retrieve data for each request,
the disks containing the FITS files for the requested region need to be opened
according to a data-placement table
containing file information and file location;
then the FITS files requested can be read.
After reading the data, the disks were closed to save energy.
During the simulation, we analyzed the number of disk-open operations,
the data-access request response time, and the overhead energy consumption.

In addition, we doubled the FITS files to evaluate the performance of AstroLayout,
and the evaluation uses the same requests.
Because the doubled data size, more disks will be used, and each disk will
store data in smaller sky areas.

\subsection{The Optimized Data Placement}

Using AstroLayout, it generated an optimized data placement SADP;
then, according to this placement, it copied FITS files into the archive,
which contained a total of 12 disks.
Under traditional data placement, 11 disks would be required.
Thus, the disk usage was 89.40\% for SADP
and 97.53\% for the traditional method.
In the case of the double size dataset, it used 23 disks under SADP, and its disk usage was 93.28\%.
However, the traditional data placement maintained a usage of 97.53\%.

\begin{figure}[ht]
  \centering
  \includegraphics{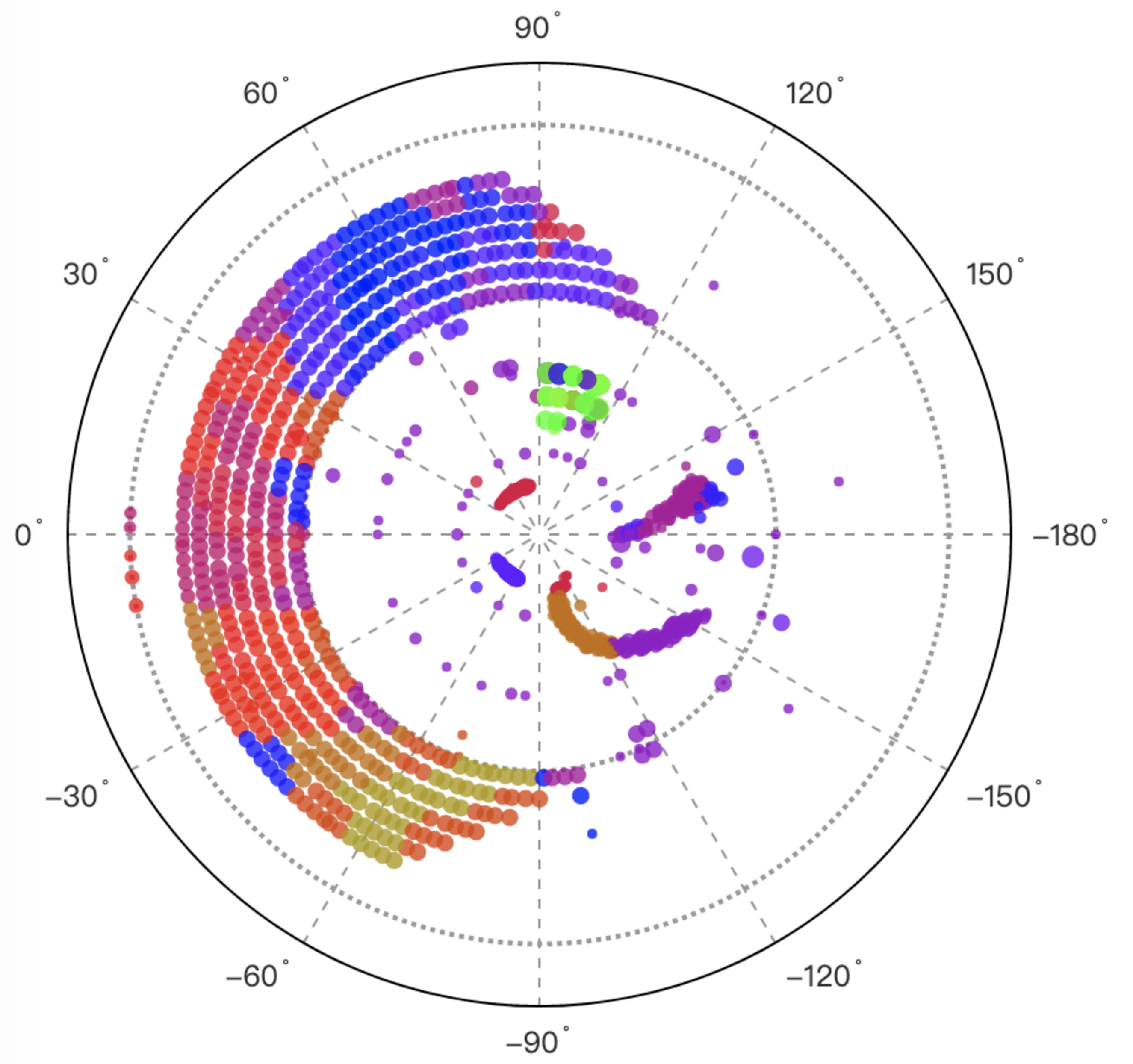}
  \caption{SADP visualization in orthographic projection.
  All data were separated into 23 groups of different colors.}
  \label{fig:fig7}
\end{figure}

We visualized SADP using the orthographic projection shown in Figure \ref{fig:fig7}.
The figure contains 23 colors,
different colors indicate that the data should be distributed into different disks,
whereas adjacent areas in the same color indicate that the data of neighboring sky areas
need to be stored in one disk.
It can be clearly seen that the data were aggregated by spatial attribute.

\subsection{Disk-Open Operations}

In long-term archives, disks are closed to save energy;
however, the disks containing the requested data need to be opened when requests arrive.
Opening and closing disks frequently will cause a significant response delay
and energy consumption.
Thus, to achieve higher performance,
the data placement should open fewer disks for the same request.

\begin{figure}[ht]
  \centering
  \includegraphics{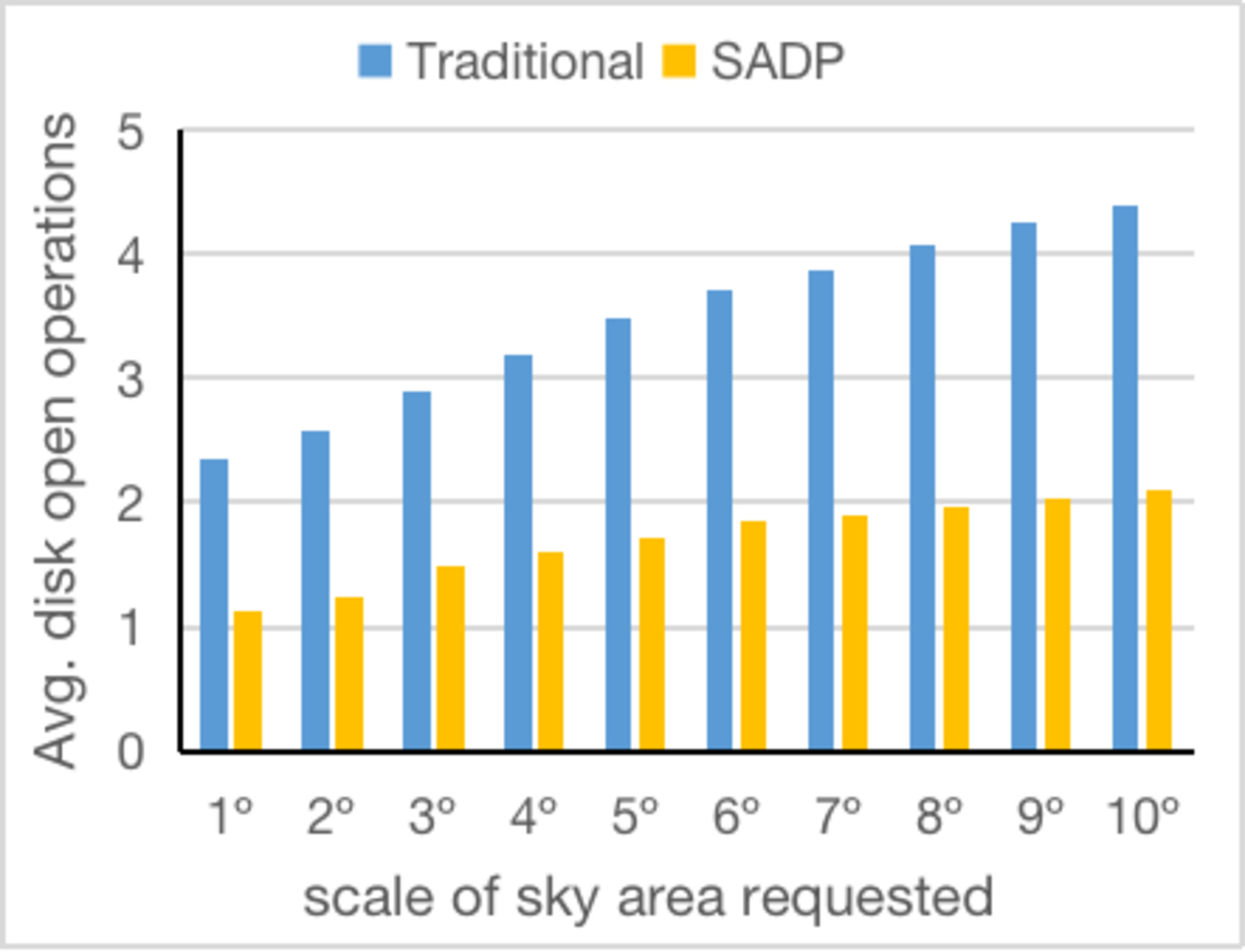}
  \caption{Comparison of disk-open operations.}
  \label{fig:fig8}
\end{figure}

\begin{figure}[ht]
  \centering
  \includegraphics{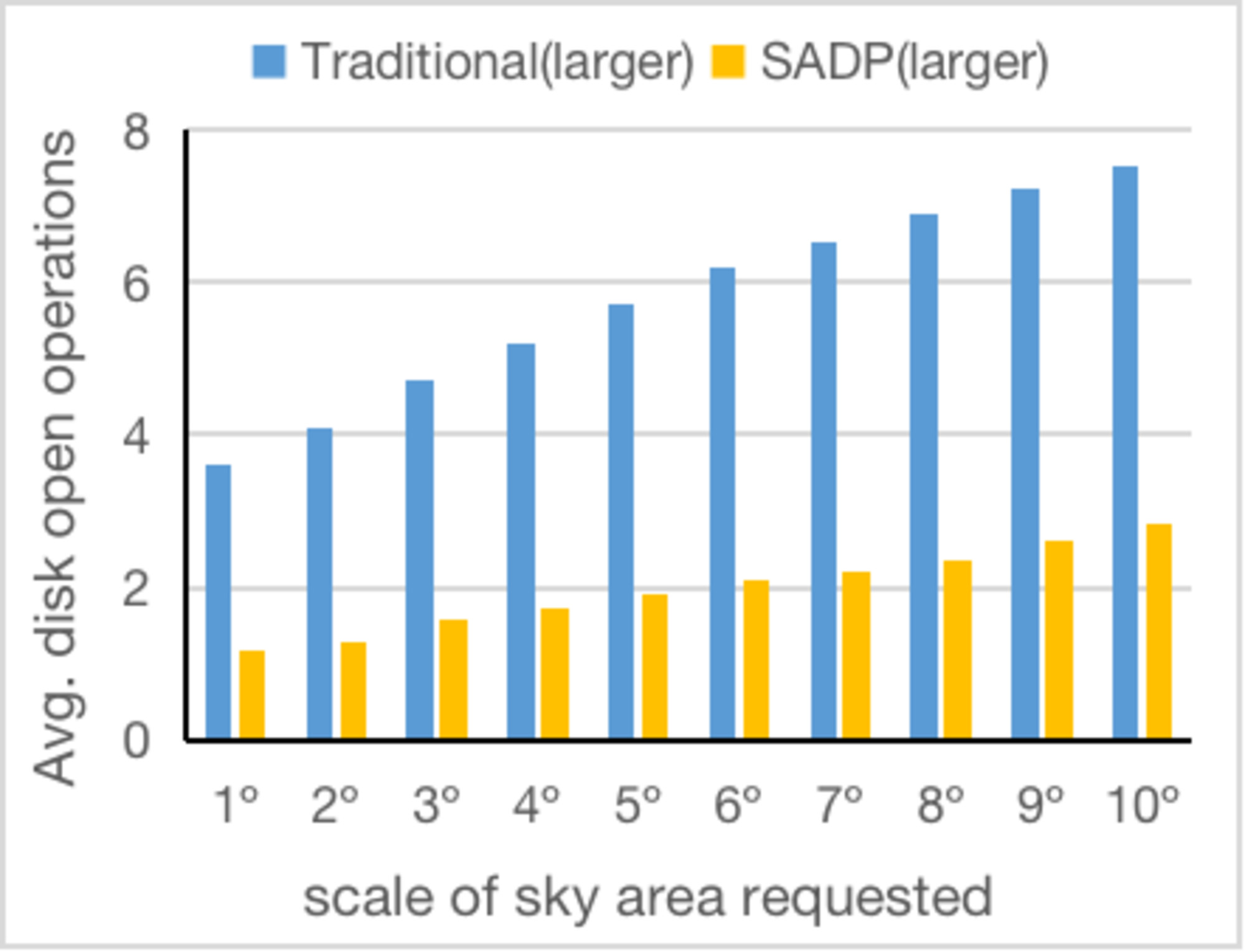}
  \caption{Comparison of disk-open operations for the double size dataset.}
  \label{fig:fig9}
\end{figure}

A comparison of disk-open operations under traditional data placement and SADP
(when processing requests in different scales) is shown in
Figures \ref{fig:fig8} and \ref{fig:fig9}.
Using the SADP reduced the number of open-disk operations
compared with traditional data placement,
and the advantages of SADP are even more significant for the double size dataset.
When the requested scale is $5^\circ$ for the double size dataset,
the number of disk-open operations under SADP is approximately 33.18\% that of traditional data placement.

\subsection{Data-Access Request Response Time}

The time taken to respond to data-access requests is an accurate representation of the performance.
Here, this time period is determined by two processes:
the time taken to read the files and the time taken to activate the disks.
The more disks to opened for the same request, the larger the response time.
Thus, data placement with aggregated data will outperform that of non-aggregated data.

\begin{figure}[ht]
  \centering
  \includegraphics{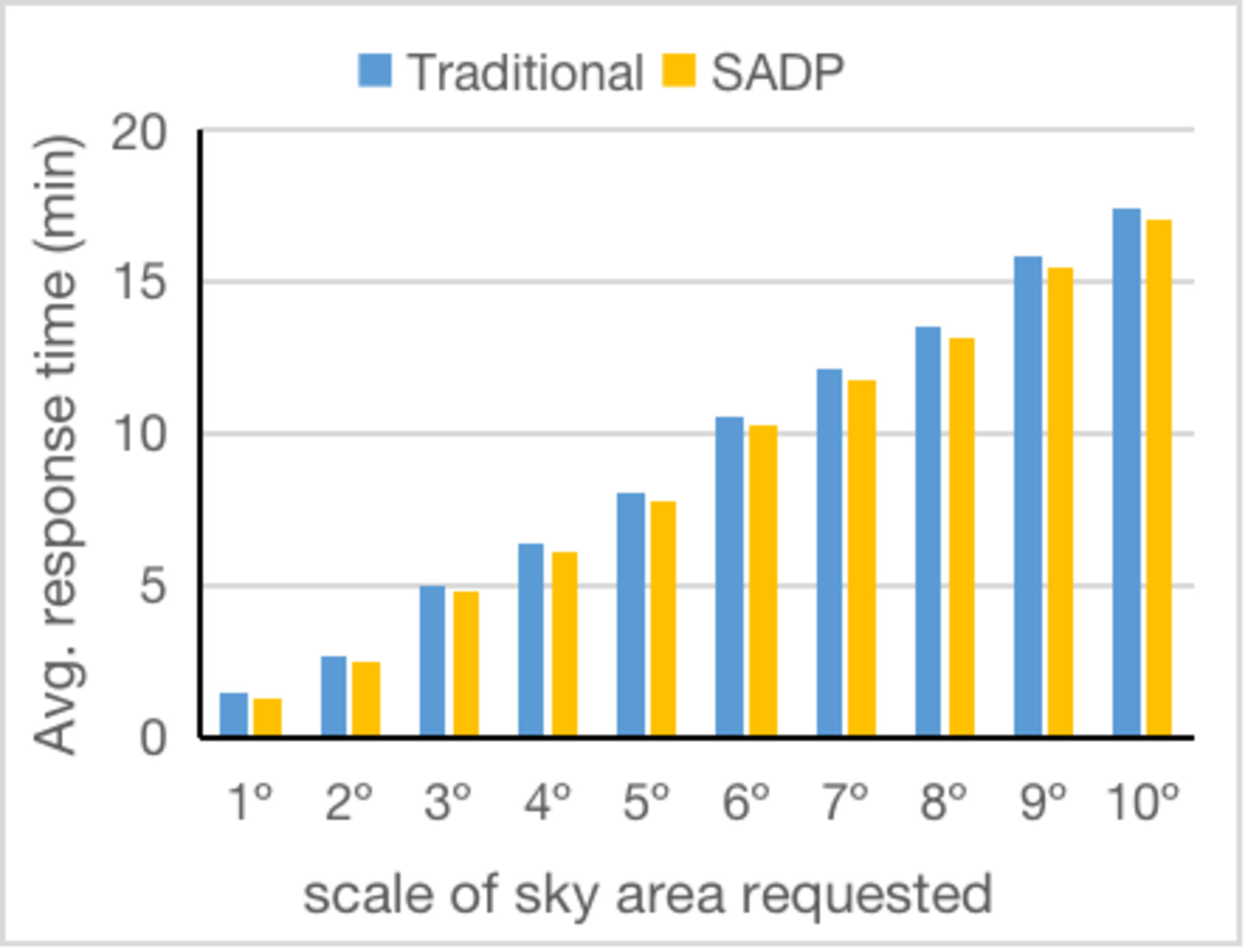}
  \caption{Comparison of the data-access request response times.}
  \label{fig:fig10}
\end{figure}

\begin{figure}[ht]
  \centering
  \includegraphics{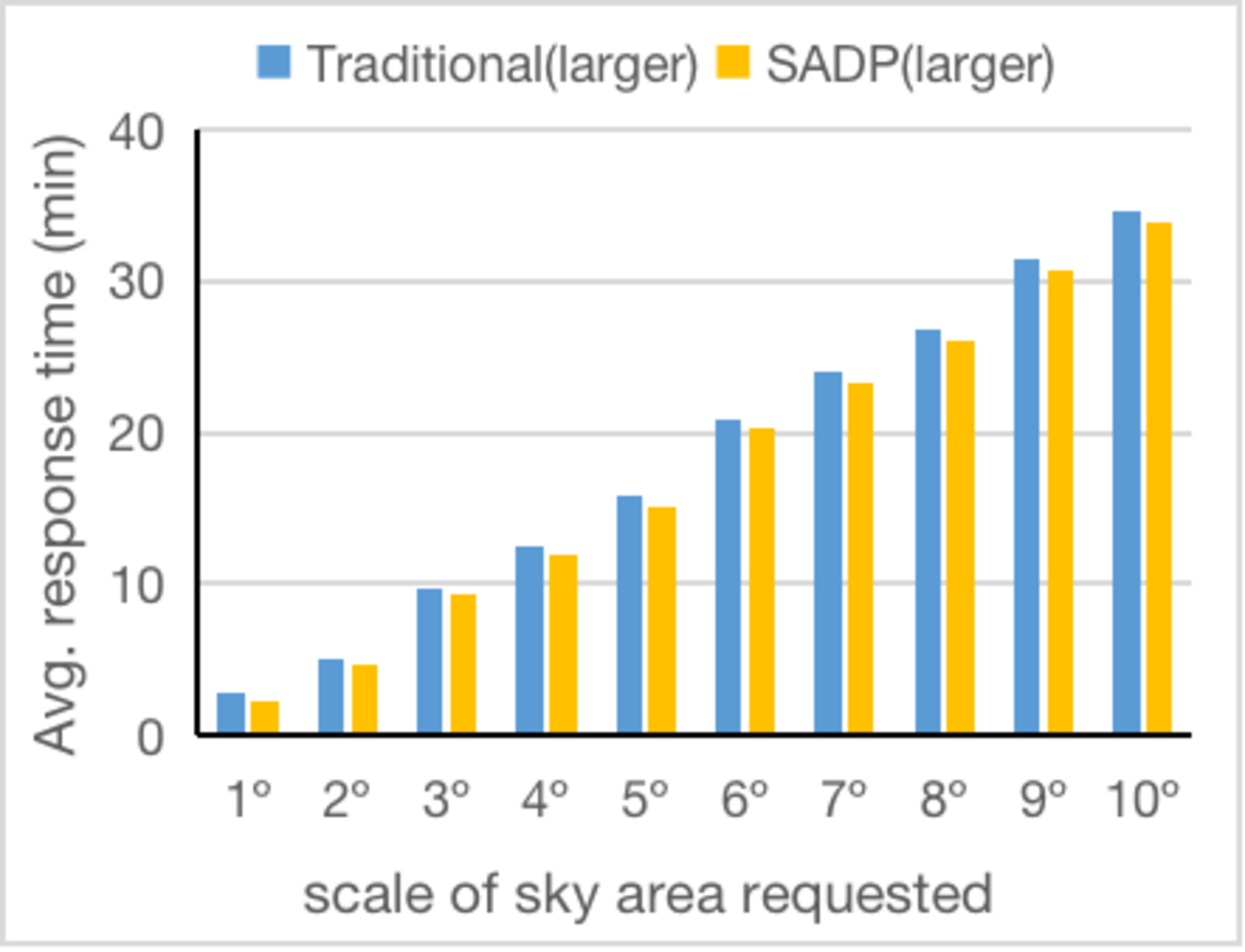}
  \caption{Comparison of the data-access request response times for the double size dataset.}
  \label{fig:fig11}
\end{figure}

The comparison of data-access request response times between traditional data placement and SADP
is shown in Figures \ref{fig:fig10} and \ref{fig:fig11}.
Using the SADP will reduces the number of open-disk operations;
thus, it also reduces the response time compared with traditional data placement.
When the request scale is $2^\circ$ for the double size dataset,
the response time using SADP is about 90.86\% that of the traditional data placement.

\subsection{Overhead Energy Consumption}

The energy consumption of data-reading processes is identical across different data placement methods
because the number of FITS files requested for one query is fixed.
The overhead energy consumption represents the energy consumption when reading costs are excluded.

\begin{figure}[ht]
  \centering
  \includegraphics{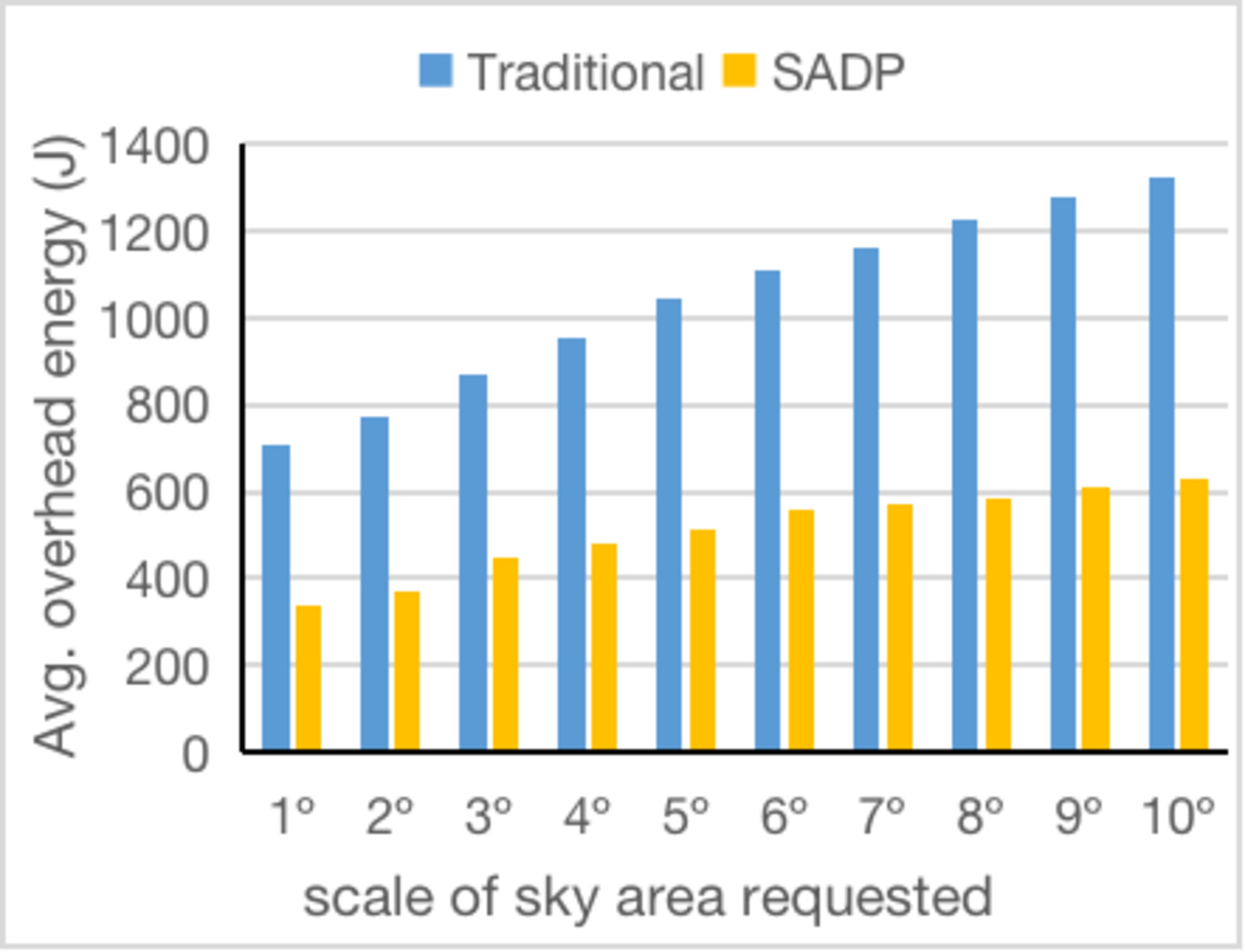}
  \caption{Comparison of the overhead energy consumption.}
  \label{fig:fig12}
\end{figure}

\begin{figure}[ht]
  \centering
  \includegraphics{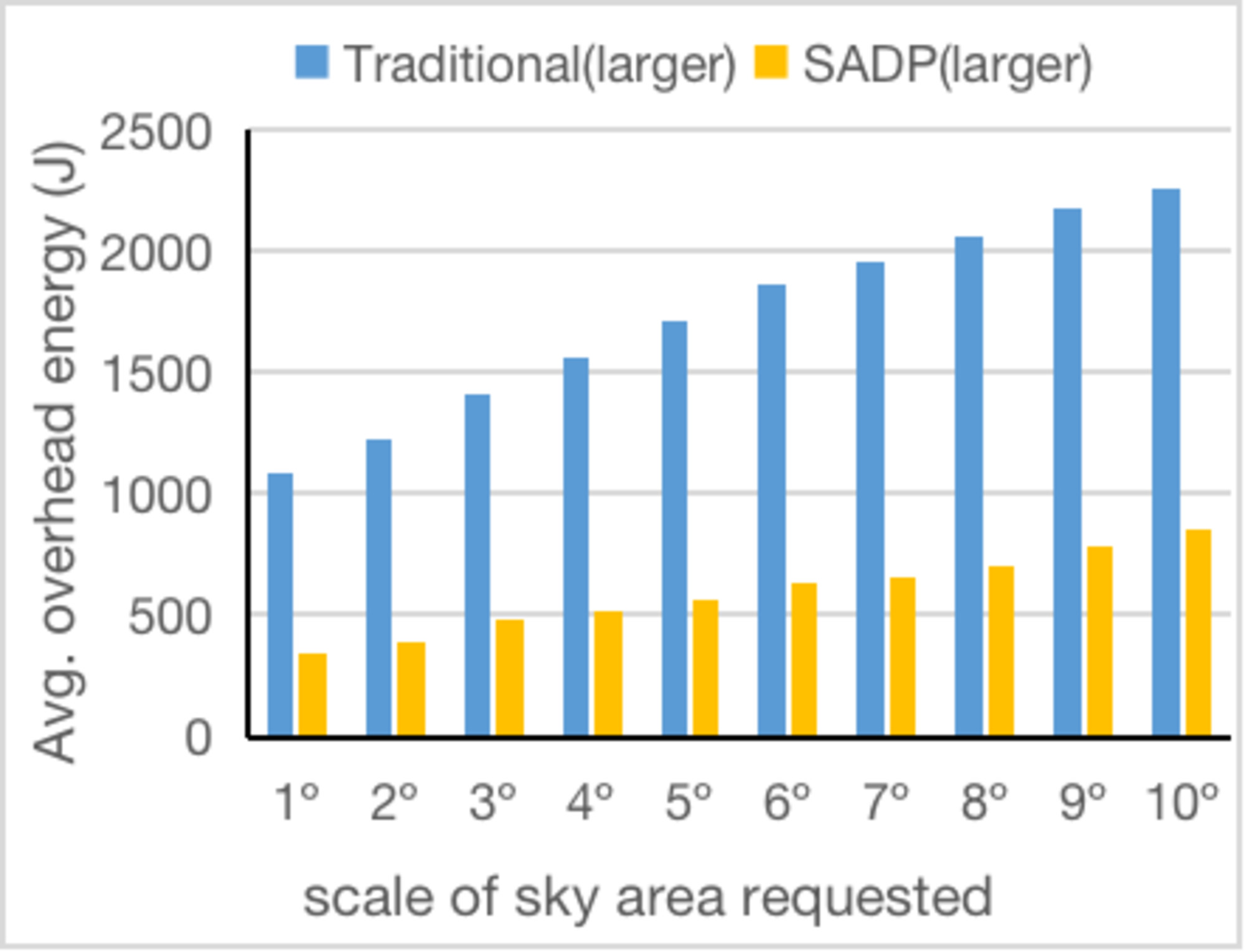}
  \caption{Comparison of the overhead energy consumption for the double size dataset.}
  \label{fig:fig13}
\end{figure}

Figures \ref{fig:fig12} and \ref{fig:fig13} show the comparison of
the overhead energy consumption using traditional data placement and SADP.
The results are similar to those of the disk-open operations,
because the overhead energy consumption is in approximately proportional to the number of disk-open operations.
SADP reduces the overhead energy consumption by more than a half
compared to traditional data placement;
thus, it is more energy efficient for double size datasets.

\section{Conclusion}

Low-cost storage is used for long-term observation data preservation,
to minimize the increasing costs.
However, data-access requests often focus on a specific observation region,
and the requested data are often located across multiple devices in the storage.
Thus, the performance and energy consumption of data-accessing procedures is a serious problem.

In this study, we developed a tool referred to as AstroLayout;
it can redistribute long-term archives of astronomical observation data
according to their spatial attributes.
AstroLayout generates a spatially aggregated data placement of the historical observation data
and copies the FITS files to archive storage using this generated placement.

According to our evaluation,
the storage system implementing SADP can respond to time-domain astronomy data-access requests
using fewer activated storage devices
compared to the traditional data-placement technique.
The results also show that AstroLayout can reduce the time taken
to respond to data-access requests and the energy consumption of the storage system.
Especially, it must be more efficient for storage realized by tapes or optical disks than that of hard disks,
owning to the device-mounting and data-accessing cost more time and energy.

The redistributed dataset is not to completely replace the original dataset.
They can act as the backup of each other.
When the redistributed dataset is partially damaged, it needs to access the original dataset to recovery.
Conversely, when a problem occurs in the original dataset,
it can also be recovered from the redistributed dataset.

In future, we will integrate AstroLayout into the Chinese Virtual Observatory (China-VO)
and conduct data archive management of the China National Astronomical Data Center (NADC).
Moreover, to enhance the program’s range of application,
we will develop the program’s compatibility with more optional features,
including the world coordinate system (WCS),
the spherical surface partition methods,
and data-aggregation methods.

\section*{Acknowledgements}
Funding: This work was supported by the Joint Research Fund in Astronomy [grant numbers U1731243, U1931130, U1731125] under the cooperative agreement of the National Natural Science Foundation of China (NSFC) and Chinese Academy of Sciences (CAS); as well as the National Natural Science Foundation of China [grant numbers 11803022, 11573019].

Data resources were supported by the China National Astronomical Data Center (NADC) and the Chinese Virtual Observatory (China-VO).

\bibliographystyle{elsarticle-harv}

\bibliography{relayout}

\end{document}